\documentclass[finallayout,an,fleqn]{w-art}
\usepackage{times}
\usepackage{w-thm}
\usepackage{isolatin1}
\theoremstyle{plain}

\theoremstyle{definition}

\usepackage[]{graphicx}
\chardef\bslash=`\\ 

\begin{document}
\DOIsuffix{theDOIsuffix}
\Volume{324}
\Issue{S1}
\Copyrightissue{S1}
\Month{01}
\Year{2003}
\pagespan{3}{}
\Receiveddate{15 November 2002}
\Reviseddate{30 November 2002}
\Accepteddate{2 December 2002}
\Dateposted{3 December 2002}



\title[Dynamics in central arcsecond]{The Galactic Center stellar
  cluster: The central arcsecond\footnote{Based on observations at
  the Very Large Telescope (VLT) of the European Southern Observatory
  (ESO) on Paranal in Chile}}


\author[R. Schödel]{R. Schödel\footnote{Corresponding author: e-mail:
     {\sf rainer@mpe.mpg.de}, Phone: +49\,89\,30000\,3837, Fax:
     +49\,89\,30000\,3490}\inst{1}}
     \address[\inst{1}]{Max-Planck-Institut f\"ur extraterrestrische
     Physik, Giessenbachstra\ss e, Garching, Germany}
\author[R. Genzel]{R. Genzel\inst{1,2}}
\address[\inst{2}]{Department of Physics, University of California,
    Berkeley, CA 94720}
\author[T. Ott]{T. Ott\inst{1}}
\author[A. Eckart]{A. Eckart\inst{3}}
\address[\inst{3}]{I.Physikalisches Institut, Universit\"at zu K\"oln,
Z\"ulpicher Stra\ss e , K\"oln, Germany}
\begin{abstract}
With 10 years of high-resolution imaging data now available on the
stellar cluster in the Galactic Center, we analyze the dynamics of
the stars at projected distances $\leq1.2''$ from the central black
hole candidate Sagittarius A* (Sgr A*).  We find evidence for radial
anisotropy of the cluster of stars surrounding Sgr A*. We
confirm/find accelerated motion for 6 stars, with 4 stars having
passed the pericenter of their orbits during the observed time
span. We calculated/constrained the orbital parameters of these
stars. All orbits have moderate to high eccentricities. The center of
acceleration coincides with the radio position of Sgr A*. From the
orbit of the star S2, the currently most tightly constrained one, we
determine the mass of Sgr A* to $3.3\pm0.7\times10^{6}$M$_{\odot}$ and
its position to $2.0\pm2.4$~mas East and $2.7\pm4.5$~mas South of the
nominal radio position.  The data provide compelling evidence that Sgr
A* is a single supermassive black hole.
\end{abstract}
\maketitle                   





\section{Introduction}

Because of its proximity the center of the Milky Way offers the unique
opportunity to study phenomena in detail (on scales $\ll$1~pc) that are
generally thought to occur in galactic nuclei. With the enigmatic
radio and X-ray source Sagittarius A* (Sgr A*), our galactic center
(GC) harbours a prime candidate for a supermassive black hole (for
reviews on the GC see e.g. Genzel, Hollenbach, and Townes 1994;
Morris and Serabyn 1996; Melia and Falcke 2001).

Near-infrared (NIR) imaging observations with speckle or adaptive
optics (AO) techniques allow resolving the central stellar cluster
with subarcsecond resolution. Since 1991, such observations were
regularly carried out with the MPE SHARP (Hofmann et al. 1995) NIR
speckle camera at the ESO NTT in La Silla, Chile (e.g. Eckart et al.
1992; Eckart et al. 1995). Stars are ideal test particles for
measuring the gravitational potential because they are not subject to
forces such as winds or magnetic fields.  First proper motion
measurements on the stars in the central few arcseconds by Eckart and
Genzel (1997) showed that the gravitational potential in the GC is
indeed dominated by a point mass of $\sim$3 million solar
masses. These results were confirmed with an independent proper motion
study by Ghez et al. (1998), who used the 10m-class Keck
telescope. Subsequently, the first accelerations of stars near Sgr A*
were found by Ghez et al. (2000) and Eckart et al. (2002). These
observations set even tighter demands on the density of the central
dark mass than the previously measured velocity dispersions and
allowed to constrain the position of the dark mass via the projected
acceleration vectors.

Here we report on an extensive analysis of the proper motions of stars
within $\sim1.2''$ of Sgr~A*, based on ten years of high resolution
observations. We put special emphasis on the bright (K$\sim$14) star
S2 close to Sgr~A*. Observations made in spring/summer 2002 during
commissioning and science verification of the new NIR camera and
adaptive optics system CONICA/NAOS (``NACO'') at the ESO VLT on El
Paranal, Chile, allowed determining a unique keplerian orbit for that
star. The orbital parameters put very tight constraints on the
position and concentration of the central dark mass.

\section{Observations and data reduction}

In spring 2002, the new NIR camera and adaptive optics system
CONICA/NAOS (``NACO'') was commissioned at the unit telescope 4 (Yepun)
of the ESO 8m-class VLT on El Paranal, Chile. With its unique
near-infrared wavefront sensor, this instrument is ideally suited for
adaptive optics observations of the Galactic Center. While optical
wavefront sensing can only be performed on a relatively weak guiding
star $\sim30''$ away from Sgr A*, the bright $K\sim6.5$mag
supergiant IRS 7 at $\sim5.5''$ from Sgr A* can be used in a
straightforward manner for wavefront corrections with NAOS. 

Reaching Strehl ratios of up to 50\% the NACO commissioning/science
verification observations of the GC provided the deepest images of
that region up to now. They also brought two extremely valuable contribitions
to our proper motion program: First, the large field of view of CONICA
enabled us to use 7 SiO maser stars for establishing an accurate
astrometry relative to Sgr A* (Reid et al. 2003). Second, the
observations covered the pericenter passage of the star S2 around Sgr
A* with a tightly sampled time series.

For the present work we compiled GC observations from three
different data sets: The NACO commissioning/science verification data
from 2002, the Gemini North observatory Galactic Center Demonstration
Science Data Set from the year 2000, and observations with the
MPE-built NIR speckle imaging camera SHARP at the ESO NTT in La
Silla, Chile, carried out between 1992 and 2002 (e.g. Eckart et al.
1995; Eckart et al. 2002).
 
Standard data reduction procedures, i.e. sky subtraction, dead/bad
pixel masking, and flat-fielding, were applied to all the imaging
data. From the SHARP speckle imaging data we selected by a mixed
automatic/manual process a few hundred frames of highest quality
for each epoch.  Selection criterion was that the first diffraction
ring around the dominant speckles of the brightest stars must be
clearly visible in the speckle frames.  Combining these frames of
highest quality resulted in SSA images with Strehl ratios $>30\%$.

Iterative blind deconvolution (IBD, Jeffries and Christou 1993) was
applied to the selected SHARP imaging data. As implementation of
IBD, we used  the publicly available IDAC program code, developed
at Steward Observatory by Matt Chesalka and Keith Hege (based on the
earlier Fortran Blind Deconvolution code - IDA - developed by Stuart
Jefferies and Julian Christou). Gemini and NACO images were
deconvolved with a Lucy-Richardson deconvolution. We obtained our
final maps after restoring the deconvolved images with a beam of
$\sim$100~mas (for SHARP and Gemini images) and $\sim60$~mas (for NACO
images) FWHM. For more details on data reduction and observations see
Schödel et al. (2003) and Genzel et al. (2003).

\section{Astrometry and proper motions}

Reid et al. (2003) combined several epochs of VLA/VLBA data to measure
the positions and proper motions of SiO maser stars in the GC. They
determined the position of Sgr~A* in NIR NACO images with an accuracy
of 10~mas. Ott et al. (2003) used their results to obtain precise
positions and proper motions for $\sim$1000 stars within $\sim$10$''$
of Sgr~A*. For the present work we established the astrometry via the
positions and proper motions of 9 sources from the Ott et al. (2003)
list. Errors on the stellar positions were determined by quadratically
combining the error from measuring the (pixel) position of the stars
in our maps with the error from the transformation into the radio
astrometric system.

We determined proper motions by linear least square fits to the time
series of stellar positions. Errors on the measured projected stellar
velocities were generally $<15\%$. We found 6 stars with significant
acceleration (deviation of $>3\sigma$ from linear motion). Three of
them, S1, S2, and S8 are well known from previous publications (Ghez
et al. 2000; Eckart et al. 2002; Schödel et al, 2002). The other three
sources, S12, S13, and S14, are fainter ($K\geq$15) sources, the
proper motions of which could only be disentangled from the high
confusion central stellar cluster with a sufficiently large data base
such as presented here (see also Schoedel et al. 2003).

\section{Radial anisotropy}

We examined the Sgr A* stellar cluster using $\gamma_{TR} =
(v_{T}^2-v_{R}^2)/v^2$ as anisotropy estimator, where $v$ is the
proper motion velocity of a star, and $v_{T}$ and $v_{R}$ its
projected tangential and radial components. A value of $+1$ signifies
projected tangential motion, $-1$ projected radial motion of a
star. The properties of the anisotropy parameter $\gamma_{TR}$ are
discussed in detail in Genzel et al. (2000). They show that an
intrinsic three-dimensional radial/tangential anisotropy is reflected
in the properties of the two-dimensional anisotropy estimator
$\gamma_{TR}$.

\begin{figure}
\begin{center}
\includegraphics[width=6cm,angle=270]{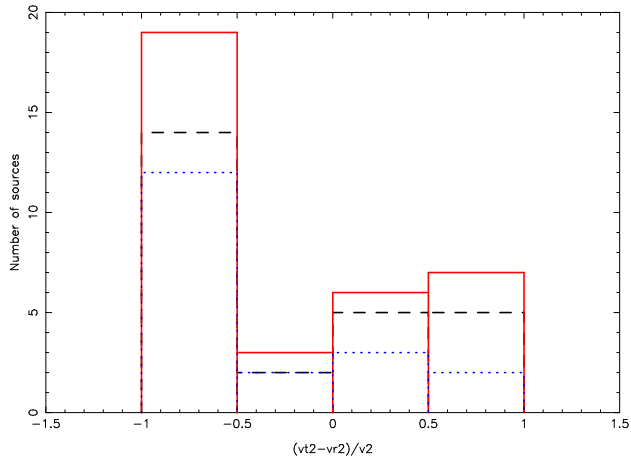}
\caption{Histogram of the anisotropy parameter
  $\gamma_{TR}=(v_{T}^2-v_{R}^2)/v^2$ of the stars in our sample (for
  the epoch 2002.7).  $v$ is the proper motion velocity; $v_{T}$ and
  $v_{R}$ are its radial and tangential components relative to
  Sgr~A*. Blue dotted lines: Stars at projected distances $<0.6$ from
  Sgr A*; black dashed lines: stars at projected distances $< 1''$ ;
  red lines: stars at projected distances $< 1.2''$
  \label{anisohist}}
\end{center}
\end{figure}

In Figure~\ref{anisohist} we show a histogram of the parameter
$\gamma_{TR}$ for different sub-samples (distance to Sgr A*
$\leq0.6''$, $\leq1.0''$, and $\leq1.2''$) of our proper motion data
for the epoch 2002.7. The projected velocities of the accelerated
stars at this epoch were estimated by linear fits to sufficiently
short parts of their trajectories.  Repeating the anisotropy analysis
for the epoch 1995.5 (where some of the stars had significantly
different positions and velocities) does not change the distribution
of counts in the histrograms significantly.  The number of stars on
projected radial orbits is $2-3\sigma$ above the number of stars on
projected tangential orbits (taking Poisson errors). The number of
stars on projected tangential orbits decreases significantly with
decreasing distance to Sgr A*.

More proper motion data are needed in order to settle the question of
anisotropy, but the present analysis presents a very intriguing
result.  Should the radial anisotropy of the Sgr A* cluster indeed be
proven to be true with the larger proper motion samples expected from
future observations, theoretical and modeling efforts will be needed
to understand this property of the Sgr A* stellar cluster.  As a
bottom line, we want to point out that the general distribution of the
anisotropy parameter definitely excludes a tangentially anisotropic
cluster. A significant tangential anisotropy would be expected in
systems with a binary black hole, where stars on radial orbits would
be ejected or destroyed preferentially (see e.g.  Gebhardt et
al. 2002).

\section{Stellar orbits}

\begin{center}
\begin{figure}
\includegraphics[width=\textwidth]{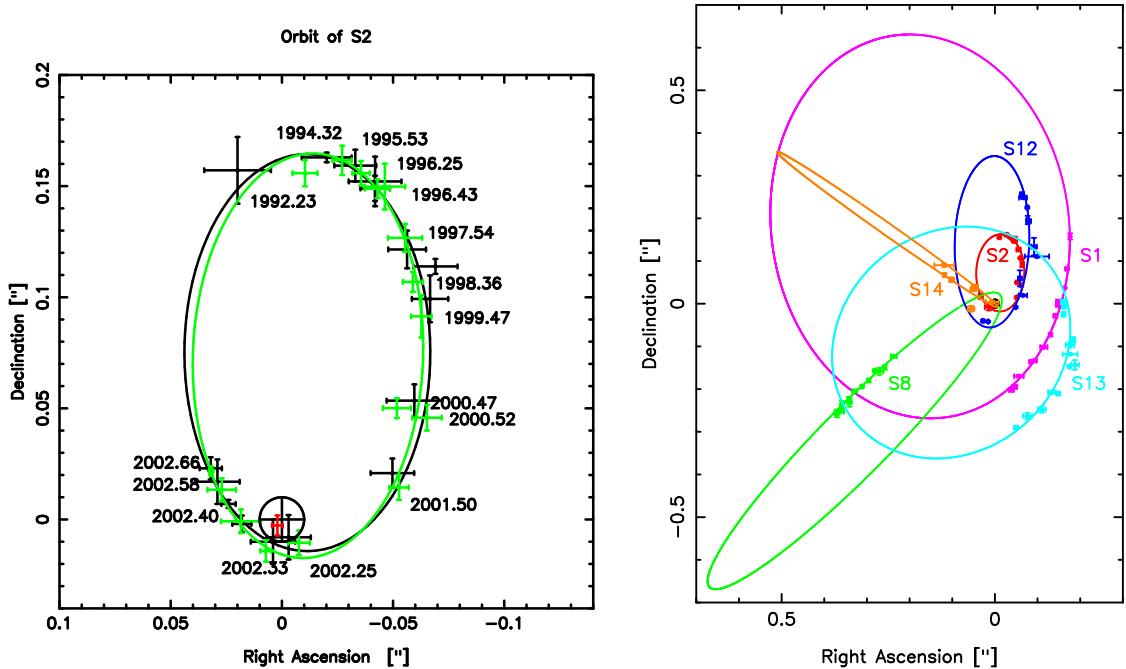}
\caption{Left panel: The orbit of S2 (black) as determined by Schödel
  et al. (2002) compared with the orbit as determined in the present
  work (green). The black cross and circle denote the radio position
  of Sgr~A* and its errors. The red cross designates the position of
  the focus of the orbit and its error resulting from the present
  analysis. Right panel: Currently, we can determine/constrain the
  orbits of 6 stars near Sgr~A*: S1, S2, S8, S12, S13, and S14. all
  orbits have moderate to high eccentricities.
  \label{orbits}}
\end{figure}
\end{center}

The NACO GC observations in spring/summer 2002 covered the pericenter
passage of the star S2 around Sgr~A* in a tightly sampled time
series. Combining these observations with SHARP imaging data since
1992 (taken from Ott et al. 2003), Schoedel et al. (2002) determined
a unique keplerian orbit for S2. In the left panel of Figure~\ref{orbits} we
compare the orbit of S2 of Schoedel et al. (2002) with the orbit of S2
as determined in the present work. There are three important
differences between the two analyses (see Schödel et al. 2003): (1)
Here, the SHARP positions were obtained with different data reduction
and analysis techniques (from a comparison with Ott et al. (2003) we
estimated an overall systematic error of $\sim$3~mas). (2) Schoedel et
al. (2002) measured the positions of S2 from one final shift-and-add
image for each NACO epoch and estimated the errors
conservatively. Here, the S2 position for each NACO observing epoch
results from measurements on several tens of individual short-exposure
NACO images, with the standard deviation taken as error. (3) We
treated the projected position of the focus of the elliptical orbit as
a free parameter in the fit (see Schödel et al. 2003).

The two analyses compare very well, with the determined orbital
parameters agreeing within the errors. In our present analysis, we
obtain a central mass of $3.3\pm0.7\times10^{6}$M$_{\odot}$. The
position of the acceleration center is offset a mere $2.0\pm2.4$~mas
East and $2.7\pm4.5$~mas South of the nominal radio position of
Sgr~A*, i.e. clearly within the error circle of the radio measurement.
This strongly supports the assumption that the dark mass is indeed
coincident with Sgr~A*. The orbit has an eccentricity of
$0.87\pm0.02$, an inclination of $45.7\pm2.6$ degrees, a period of
$15.7\pm0.74$ years, a semi-major axis of $4.54\pm0.27$~mpc, and a
pericenter distance of $0.59\pm0.10$~mpc.

Significant sections of the orbits were observed as well in the case
of the stars S12 (pericenter passage in 1995.3) and S14 (pericenter
passage in 1999.9). However, the constraints on their orbits from our
data are not very tight. S14 is identical with the source S0-16 of
A. Ghez et al. (priv. comm.), who first determined an orbit for this
source. S14 is on an extremely eccentric ($e=0.97\pm0.05$) and highly
inclined ($i>80\deg$) orbit and approaches Sgr~A* to within
$\sim$0.4~mpc (S2: 0.6~mpc). In principle, its orbit would allow to
constrain the central mass distribution even tighter than
S2. Unfortunately, the uncertainty in the orbital parameters of S14
resulting from our data is too high for this purpose.

The orbital segments observed for the stars S1 (pericenter passage
around 1999/2000), S8, and S13 are too small for determining a unique
set of parameters for them, but we constrained them by using fixed
values for the inclination angle (see Schoedel et
al. 2003). Approximate values of the inclination of the orbital planes
could be estimated from the measured acceleration of the stars and the
well known mass of the central dark object (see Figure~3). We plot all
analyzed six orbits in the right hand panel of
Figure~\ref{orbits}. All analyzed orbits have moderate to high
eccentricities. Future measurements of orbital eccentricities of more
stars near Sgr~A* will allow testing for  anisotropy of the
central cluster (see Schödel et al. 2003).

\section{Nature of the enclosed mass}

With the measured proper motions within $1.2''$ of Sgr A* we
calculated Leonard-Merritt (LM, Leonard and Merritt 1989) estimates of
the enclosed mass. In order to take the strongly variable velocity of
the 6 stars with significant acceleration into account, we produced various
velocity lists for the analysis, where we estimated the projected
velocity of these stars at different epochs. From the diffrent
lists, we obtain an average LM mass of
$3.4\pm0.5\times10^{6}$M$_{\odot}$ (for details see Schödel et
al. 2003). This agrees well with the mass estimate from the orbit of
S2.

\begin{figure}
\begin{center}
\includegraphics[width=12cm,angle=270]{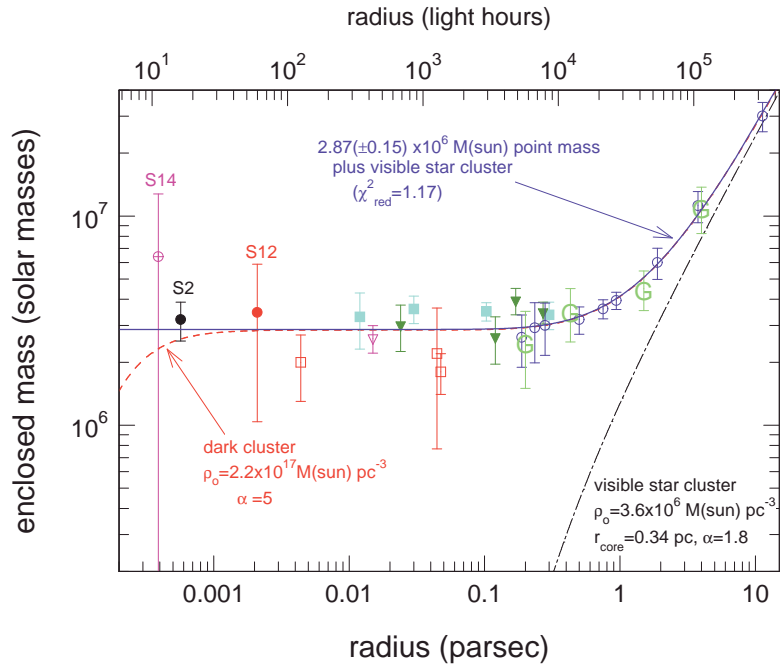}
\caption{Mass distribution in the Galactic Center assuming an 8 kpc
distance (Reid et al. 2003). The filled black circle denotes the mass
derived from the orbit of S2, the red filled circle the mass derived
from the orbit of S12, and the purple circle the mass derived from the
orbit of S14. Filled dark green triangles denote Leonard-Merritt
projected mass estimators from the present work (at 0.025~pc) and from
a new NTT proper motion data set by Ott et al. (2003), separating late
and early type stars, and correcting for the volume bias determined
from Monte Carlo modeling of theoretical clusters and assuming a
central density profile with a power-law slope of $\alpha=1.37$
(Genzel et al. 2003). An open down-pointing triangle denotes the
Bahcall-Tremaine mass estimate obtained from Keck proper motions (Ghez
et al. 1998). Light-blue, filled rectangles are mass estimates from a
parameterized Jeans-equation model, including anisotropy and
distinguishing between late and early type stars (Genzel et al.
2000). Open circles are mass estimates from a parameterized
Jeans-equation model of the radial velocities of late type stars,
assuming isotropy (Genzel et al. 1996). Open red rectangles denote
mass estimates from a non-parametric, maximum likelihood model,
assuming isotropy and combining late and early type stars (Chakrabarty
and Saha 2001). The different statistical estimates (in part using the
same or similar data) agree within their uncertainties but the
variations show the sensitivity to the input assumptions. In contrast,
the orbital technique for S2/S12 and S14 is much simpler and less
affected by the assumptions. Green letter ``G'' points denote mass
estimates obtained from Doppler motions of gas (Genzel and Townes
1987). The blue continuous curve is the overall best fit model to all
data. It is the sum of a $2.87\pm0.15\times10^{6}$ M$_{\odot}$ point
mass, plus the visible outer stellar cluster of central density
$3.6\times10^{6}$M$_{\odot}$pc$^{-3}$, core radius 0.34 pc and
power-law index $\alpha=1.8$. The grey long dash-short dash curve
shows the same stellar cluster separately, but for a infinitely small
core (i.e. a 'cusp'). The red dashed curve is the sum of the stellar
cluster, plus a Plummer model of a hypothetical very compact (core
radius $\sim$0.00019~pc) dark cluster of central density
$2.2\times10^{17}$M$_{\odot}$pc$^{-3}$.
\label{encmass}}
\end{center}
\end{figure}

Figure~\ref{encmass} is a plot of the measured enclosed mass against
distance from Sgr A*, in close analogy to Figure~17 of Genzel et
al. (2000) and Figure~3 of Schödel et al. (2002).  The data show that
the central mass distribution is remarkably well described by the
potential of a point mass over 3 orders of magnitude in spatial scale,
from 0.8 light days to 2 light years. The contribution of the extended
stellar cluster around SgrA* to the total mass cannot be more than
mostly a few hundred solar masses within the peri-center distance of
S2 (Mouawad et al. 2003). Fitting a model composed of a point mass
plus the visible outer stellar cluster with a core radius of 0.34~pc and a
power-law slope of $\alpha=1.8$ to the data gives a value of
$2.9\pm0.2\times10^{6}$M$_{\odot}$ for the central dark mass. This
agrees within the errors with the LM mass estimate of the innermost
stars and with the masses calculated from the orbital parameters of S2
and S12. It is higher than the $2.6\pm0.2\times10^{6}$M$_{\odot}$
given by Schödel et al. (2002), but the two values agree within their
errors. The main differences of the present analysis to Schödel et
al. (2002) are: (1) The error of the mass estimate from the orbit of
S2 has been reduced by taking the position of the orbital focus
explicitly into account. (2) The innermost LM mass estimate of Schödel
et al. (2002) was based on the Ott et al. (2003) data. It has been
replaced by the LM mass estimate from the present work, which is based
on a more abundant data base in the region within $\sim1''$ of Sgr
A*. (3) The LM mass estimates in Figure~3 of Schödel et al. (2002) and
Figure~17 of Genzel et al. (2000) were corrected downward by 5-10\%
because they assumed a power-law slope of $\alpha=1.8$ for the stellar
cluster in the innermost few arcseconds. Here, we use a power-law
slope of $\alpha \approx 1.4$ for the stellar cusp around Sgr~A*
(Genzel et al.  2003). This means that the LM mass estimates have
previously been underestimated by $\sim$10\% .

The orbit of S2 places very tight constraints on the distribution of
the central dark mass: If the central point mass were replaced by a
Plummer model cluster of dark astrophysical objects, its central mass
density would have to exceed 2.2x10$^{17}$M$_{\odot }$pc$^{-3}$,
almost 5 orders of magnitude greater than previous estimates (Ghez et
al. 1998, 2000; Genzel et al. 2000). The lifetime of such a
hypothetical cluster would be $<10^{5}$ years (Maoz 1998).  An
alternative model to supermassive black holes in galactic nuclei are
balls of heavy, degenerate neutrinos (Tsiklauri and Viollier 1998;
Munyaneza and Viollier 2002). In order to explain the whole mass range
of dark central objects in galaxies with such a model, the neutrino
mass  cannot be higher than $17$keV (Melia and Falcke
2001). However, the orbital parameters of S2 would demand a neutrino
mass of $>50$keV in the case of the dark mass in the Galactic Center.

The only dark particle matter explanation that cannot be ruled out by
the present data is a ball of bosons (Torres et al.
2000). However, it would be hard to understand how the bosons first
manage to reach such a high concentration, and then avoid forming a
black hole by accretion of the abundant gas and dust in the GC. We
therefore conclude that the most probable form of the dark mass at the
center of the Milky Way is a single, supermassive black hole.

\begin{acknowledgement}
We like to thank the ESO NTT team for their help and support during
ten years of observations with the SHARP guest instrument.  

We thank the NAOS and CONICA team members for their hard work, as well
as the staff of El Paranal and the Garching Data Management Division
for their support of the commissioning and science verification.

Based on observations obtained at the Gemini Observatory, which is
operated by the Association of Universities for Research in Astronomy,
Inc., under a cooperative agreement with the NSF on behalf of the
Gemini partnership: the National Science Foundation (United States),
the Particle Physics and Astronomy Research Council (United Kingdom),
the National Research Council (Canada), CONICYT (Chile), the
Australian Research Council (Australia), CNPq (Brazil) and CONICET
(Argentina). 

\end{acknowledgement}

\end{document}